\documentclass[12pt]{article}

\usepackage{newtxtext,newtxmath}
\usepackage{soul}
\usepackage{graphicx}
\usepackage{subcaption}

\usepackage[letterpaper,margin=1in]{geometry}

\renewenvironment{abstract}
	{\quotation}
	{\endquotation}

\date{}

\makeatletter
\renewcommand{\fnum@figure}{\textbf{Figure \thefigure}}
\renewcommand{\fnum@table}{\textbf{Table \thetable}}
\makeatother

\usepackage{scicite}
\usepackage{caption}
\usepackage{float}
\usepackage{url}

\def\scititle{
	Sub-terahertz Spin Relaxation Dynamics of Boron-Vacancy Centers in Hexagonal Boron Nitride
}
\title{\bfseries \boldmath \scititle}

\author{
Abhishek~Bharatbhai~Solanki$^{1}$, Yueh-Chun~Wu$^{2}$, Hamza~Ather$^{3,4}$,
Priyo~Adhikary$^{1}$, \and Aravindh~Shankar$^{1}$, Ian~Gallagher$^{2}$, Xingyu~Gao$^{3}$, Owen~M. Matthiessen$^{1,4}$, \and
Demid~Sychev$^{1,4}$, Alexei~Lagoutchev$^{4}$, Tongcang~Li$^{1,3,4,5}$, Yong~P.~Chen$^{1,3,4,5,6,7}$, \and
Vladimir~M.~Shalaev$^{1,3,4,5,\ast}$, Benjamin~Lawrie$^{2,\ast}$, Pramey~Upadhyaya$^{1,5,\ast}$ \and
\small$^{1}$Elmore Family School of Electrical and Computer Engineering, Purdue University, West Lafayette, IN, USA. \and
\small$^{2}$Materials Science and Technology Division, Oak Ridge National Laboratory, Oak Ridge, TN, USA. \and
\small$^{3}$Department of Physics and Astronomy, Purdue University, West Lafayette, IN, USA. \and
\small$^{4}$Birck Nanotechnology Center, Purdue University, West Lafayette, IN, USA. \and
\small$^{5}$Purdue Quantum Science and Engineering Institute (PQSEI), Purdue University, West Lafayette, IN, USA. \and
\small$^{6}$Institute of Physics and Astronomy and Villum Center for Hybrid Quantum Materials and Devices, \and \small Aarhus University, Aarhus, Denmark. \and
\small$^{7}$WPI-AIMR International Research Center for Materials Sciences, Tohoku University, Sendai, Japan. \and
\small$^\ast$Corresponding authors. Emails:  prameyup@purdue.edu, lawriebj@ornl.gov, shalaev@purdue.edu
}

\begin{document} 

% Insert the title and author list
\maketitle

% Abstract, in bold
% There are strict length limits, and not all formats have abstracts.
% Consult the journal instructions to authors for details.
% Do not cite any references in the abstract.

\begin{abstract} \bfseries \boldmath
Quantum sensors based on spin-defect relaxation have become powerful tools for detecting faint magnetic signals, yet their operation has remained largely confined to low magnetic fields and gigahertz frequencies. Extending such sensors into high-field ($> 0.3$ T) and sub-terahertz regimes would enable quantum metrology across a wide range of electromagnetic phenomena and scientific applications, but has proven challenging. Here, we demonstrate that negatively charged boron vacancies ($\mathrm{V_B^-}$) in two-dimensional hexagonal boron nitride can function as relaxation-based quantum sensors operating up to 0.2 terahertz. Their uniform spin-orientation and persistent spin-contrast at high fields enable direct measurement of intrinsic spin relaxation across previously unexplored temperature and frequency regimes. We also reveal a crossover in relaxation behavior \textemdash initially decreasing at low fields before rising at higher fields \textemdash consistent with the emergence of single-phonon-induced resonant noise that becomes significant at sub-terahertz frequencies. These results establish $\mathrm{V_B^-}$ centers as a versatile platform for quantum sensing in the sub-terahertz, high-field regime.
\end{abstract}

\noindent
Solid-state spin defects offer a unique combination of optically addressable spin states and long spin coherence times, making them highly attractive for quantum applications \cite{Quantumguidelines}. Quantum sensors based on the relaxation dynamics of spin-defects, most notably the nitrogen-vacancy ($\mathrm{NV^-}$) center in diamond \cite{doherty}, leverage their pronounced sensitivity to environmental magnetic noise \cite{Thiel, DemlerDecherence, DemlerBKT, Pasupathynoisespectroscopy, casola2018probing, Flebus2018_AFM, Demler2019noiseMagnetometry, AbhishekPRR, wu2025nanoscale}, enabling a broad range of scientific applications \cite{Degan}. However, the operation of these sensors is generally limited to low magnetic fields ($<0.3 $ T) \textemdash corresponding to low gigahertz (GHz) frequencies. Accessing sub-terahertz (sub-THz) spin splitting requires precise alignment of Tesla-scale magnetic fields with the defect's symmetry axis to preserve the optical spin-contrast that forms the basis of spin-state readout techniques \cite{hopper2018spin, maletinkskyHighfield,Sahin2022HighField}. At high magnetic fields, the photodynamics of the $\mathrm{NV^-}$ center become increasingly complex due to strain-dependent excited-state level anti-crossings, which lead to spin mixing \cite{maletinkskyHighfield}. Furthermore, the fundamental mechanisms governing the longitudinal ($T_1$) and transverse spin-relaxation ($T_2$) times at high magnetic fields remain poorly understood, presenting a key challenge for realizing quantum sensing at Tesla-scale fields and sub-terahertz frequencies \cite{TerahertzNV}. 

On another front, recent efforts have increasingly focused on identifying alternative spin-defect platforms \cite{rareeartherbium, GroupIVdefects, SiCspindefects, VaidyaReview}, such as ensembles of negatively charged boron-vacancy centers ($\mathrm{V_B^-}$) in van der Waals hexagonal boron nitride ($\mathrm{hBN}$) \cite{VaidyaReview}. Unlike their bulk counterparts, $\mathrm{V_B^-}$ spin defects can be engineered in few-layer $\mathrm{hBN}$ \cite{Gao2021PlasmonicEnhanced,Xu2023EnhancedEmission,JacquesthinhBN, ChunhuithinhBN}, enabling unique ultra-thin quantum sensors positioned within a few nanometer distance of the target materials. The spin-dipole moment of the $\mathrm{V_B^-}$ center is fixed along the out-of-plane direction, enabling consistent alignment of an external uniaxial magnetic field with all defects in the ensemble. Moreover, their photodynamics remain simple and predictable at high fields \cite{Jacques-field-dependenthBN}, up to several Teslas, making them particularly attractive candidates for high-field quantum sensing applications. Dense ensembles of co-aligned spin defects in few-layer $\mathrm{hBN}$ also offer a unique platform to study many-body systems, including the impact of disorder, dimensionality, and dipolar interactions on nanoscale spin-dynamics \cite{Hydrodynamics2021, Davis2023}. 

In this work, we demonstrate that negatively charged boron vacancies ($\mathrm{V_B^-}$) in $\mathrm{hBN}$ function as spin–relaxation–based quantum sensors operating in the sub-terahertz regime. Leveraging this capability, we probe the intrinsic noise mechanisms governing temperature-dependent longitudinal spin relaxation ($1/T_1$) of $\mathrm{V_B^-}$ centers across magnetic fields from $0-7$ T ($\sim 3.5$ GHz$-0.2$ THz), revealing a rich, non-monotonic behavior. At low fields ($0-0.04$ T), where most prior studies have focused, the relaxation rate decreases before saturating. In the higher-field regime ($0.17-7$ T), it first decreases up to $\sim 1.8$ T, then reverses trend and increases. This non-monotonicity persists across all temperatures ($15-250$ K) and strengthens at lower temperatures. A phenomenological model capturing the interplay between phonon-induced and Lorentzian noise (that could be assigned to spin–spin interactions) is introduced to explain these trends. Importantly, we find that the first-order single-phonon (direct) relaxation drives the upturn beyond $\sim 1.8$ T and remains active up to $\sim 100$ K, unlike the low-field case where it is relevant only at millikelvin temperatures \cite{spontaneousphonon}. At higher temperatures, a field-independent two-phonon process starts to dominate, diminishing the non-monotonicity. Notably, we also observe stretched-exponential relaxation profiles, suggesting complex dynamics dominated by disorder. Together with demonstrated sub-terahertz relaxometry, these insights establish a microscopic foundation for high-field quantum sensing across a broad range of sub-terahertz applications.

\subsection*{Experimental results}

\begin{figure}[H]
    \centering
    \includegraphics[width=1.0\linewidth]{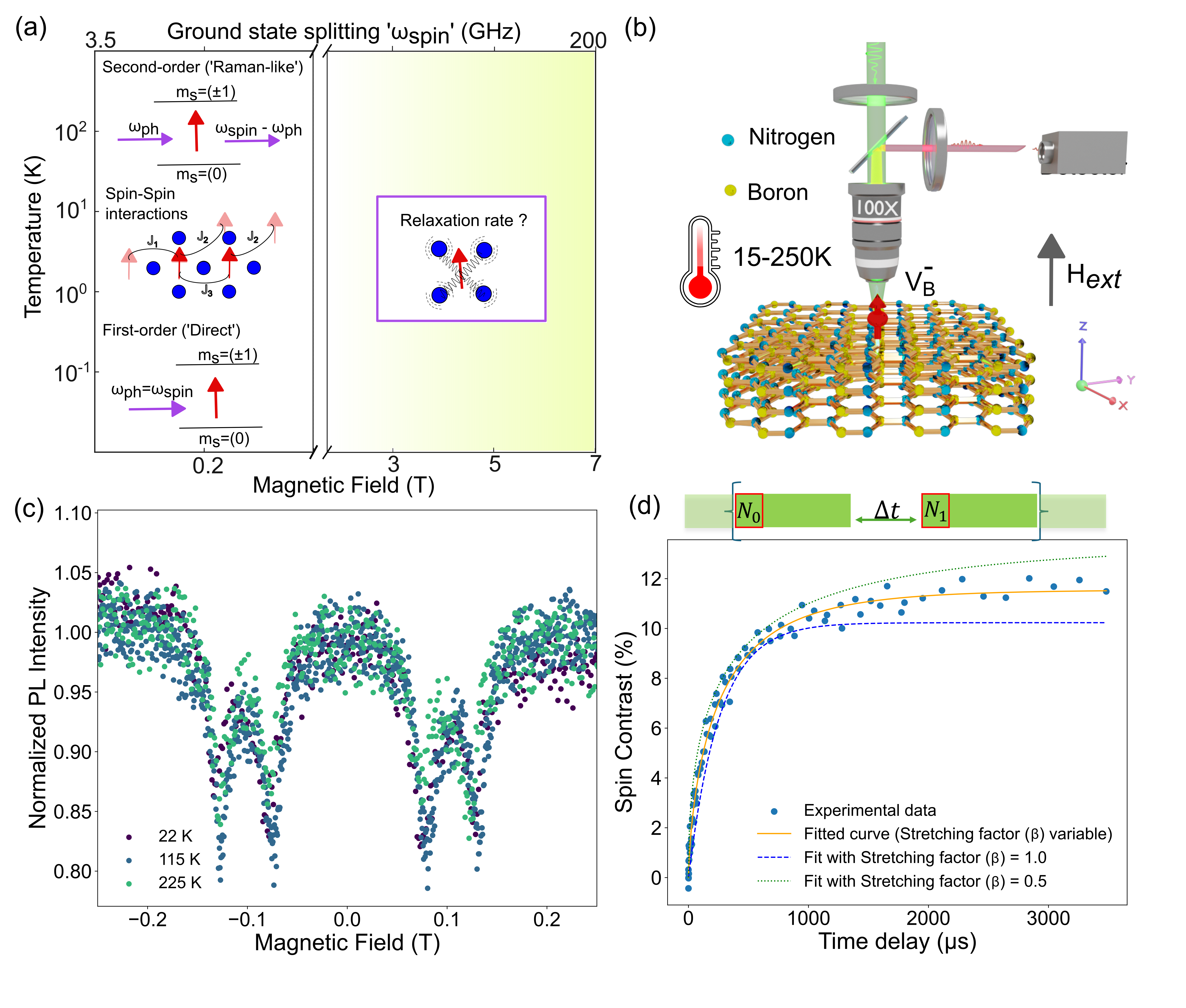}
    \caption{Spin-relaxation mechanisms and measurement scheme (a) The experimental setup comprising a $100x$ microscope objective and a layered $\mathrm{hBN}$ sample mounted on the cold-plate of a cryostat with an out-of-plane magnetic field ${H_{ext}}$. The illustrated pulse sequence shows initialization and readout pulses (green) separated by variable delay $\Delta t$; PL is measured during the interval highlighted in red.  (b) Experimental measurement of the relaxation rate for magnetic fields $H_{ext} = 0.03, 7$ T at a fixed sample temperature $T=50$ K. The experimental data is overlaid with a stretched exponential fit. Inset: Measured spin-contrast amplitude ($C_0$) as a function of external magnetic field ($H_{ext}$).  (c) Schematic diagram illustrating the key spin-relaxation processes in commonly used solid-state spin defects. The upper $x$-axis indicates the ground-state spin splitting ($\mathrm{\omega_{spin}}$) of a representative spin defect, and the lower $x$-axis shows the corresponding magnetic fields ($\sim 200$ GHz at $7$ T). A dark red arrow represents a central spin-dipole, coupled to surrounding nuclear spins (blue), other identical spin defects, and nearby paramagnetic impurities (light red) with dipolar interaction strength $J_i$. The ground state is also coupled to lattice ($\mathrm{\omega_{ph}}$) through either resonant first-order (direct) or second-order (Raman-like) spin-phonon processes. The nature of the relaxation dynamics at very high magnetic fields remains largely unexplored.} 
    \label{fig:intro}
\end{figure}

The ground-state Hamiltonian ($H_{spin}$) of $\mathrm{V_B^-}$ defects can be expressed as \cite{Gottscholl2020SpinDefects, Gottscholl2021HexagonalBN}
\begin{equation}
    H_{spin}=hD(S_z^2-S(S+1)/3)+hE(S_x^2-S_y^2)+g\mu_{B}H_{ext}S_z,
\end{equation}
where $D$ and $E$ denote the axial and transverse zero-field ground state splitting parameters respectively, $h$ is Planck's constant, $g$ is the Land\'{e}-g factor, $\mu_B$ is the Bohr Magneton, $S_{x,y,z}$ are the spin operators, $S=1$ for spin-triplet levels, and $H_{ext}$ is the external magnetic field aligned with the out-of-plane symmetry axis of the spin defect. At zero field, the ground-state splitting (ZFS) is $\sim 3.5$ GHz and reaches the sub-THz range for several Tesla applied along the quantization axis. When placed in an environment with fluctuating magnetic noise, the spin-defect's relaxation becomes a probe of the noise resonant with the ground-state splitting. To explore the potential of $\mathrm{V_B^-}$ centers as sub-THz quantum sensors, we use the experimental setup and pulse sequence shown in Fig.~\ref{fig:intro}(a) (see Methods). 

A $10 ~\mu s$ laser pulse initializes the ground state in the $m_s=0$ state, followed by a variable time delay ($\Delta t$) and a subsequent readout pulse. The $\mathrm{PL}$ signal is measured for a short duration ($1.5 ~\mu s$) at the beginning of both laser pulses. The signal during the readout pulse ($N_1$) reflects the time-dependent ground-state spin population as the system relaxes towards thermal equilibrium, while the signal measured during initialization ($N_0$) serves as the reference signal corresponding to the $m_s=0$ state. The spin-contrast as a function of time is calculated as $C=\frac{N_0-N_1}{N_0}$. Representative spin-contrast curves at low ($0.03$ T) and high fields ($7$ T) are shown in Fig.~\ref{fig:intro}(b). Phenomenologically, the spin-contrast is well described by a stretched exponential $C_0(1-\exp{{(-(\frac{\Delta t}{T_1})^\beta)}})$ \cite{Depolarization}, where $C_0$ is the spin-contrast amplitude, $\Delta t$ is the dark time interval, $T_1$ is the spin-relaxation time, and $\beta$ is the stretching factor. Crucially, $C_0$ remains large ($\sim 12 \%$) up to the highest field ($7$ T) studied here (see inset of Fig.~\ref {fig:intro}(b)), demonstrating that $\mathrm{V_B^-}$ spin defects can probe their environment's noise, as encoded in their relaxation rate ($1/T_1$), in the sub-THz range. Next, we take advantage of this capability to probe the noise environment intrinsic to $\mathrm{V_B^-}$ spin defects across $3.5$ GHz$-0.2$ THz and $15-250$ K.

Previous works have largely focused on defect spin relaxation at low fields, corresponding to an energy scale significantly smaller than that of typical acoustic phonons \cite{jarmola2012,Liu2021TemperatureShifts, hBNT1}. Consequently, the longitudinal spin-relaxation process is primarily governed by second-order spin-phonon coupling mediated by Raman-like scattering involving two high-energy phonons (Fig.~\ref{fig:intro}(c)) \cite{jarmola2012,maze_1, hBNT1}. In both $\mathrm{V_B^-}$ centers and $\mathrm{NV^-}$ centers, this mechanism leads to characteristic temperature dependence of the spin relaxation rate, exhibiting $T^2$ \cite{hBNT1} and $T^5$ \cite{jarmola2012} scaling behavior, respectively. In contrast, first-order spin-phonon coupling via direct (resonant) scattering of single phonons is relevant at millikelvin temperatures \cite{spontaneousphonon, maze_1} (Fig.~\ref{fig:intro}(c)). Dense ensembles of spin defects also exhibit additional relaxation channels arising from dipolar spin-spin interactions between defects and a rapidly fluctuating spin-bath \cite{jarmola2012,Depolarization,surfacespins} (Fig.~\ref{fig:intro}(c)). Indeed, relaxation mediated by spin-spin interaction becomes the dominant source of spin relaxation below a critical temperature, where phonon-mediated processes are strongly suppressed \cite{jarmola2012}. The goal of the remainder of the article is to address how these mechanisms interplay for $\mathrm{V_B^-}$ centers, particularly in the less studied high-field regime.

\begin{figure}[H]
    \centering
    \includegraphics[width=\linewidth]{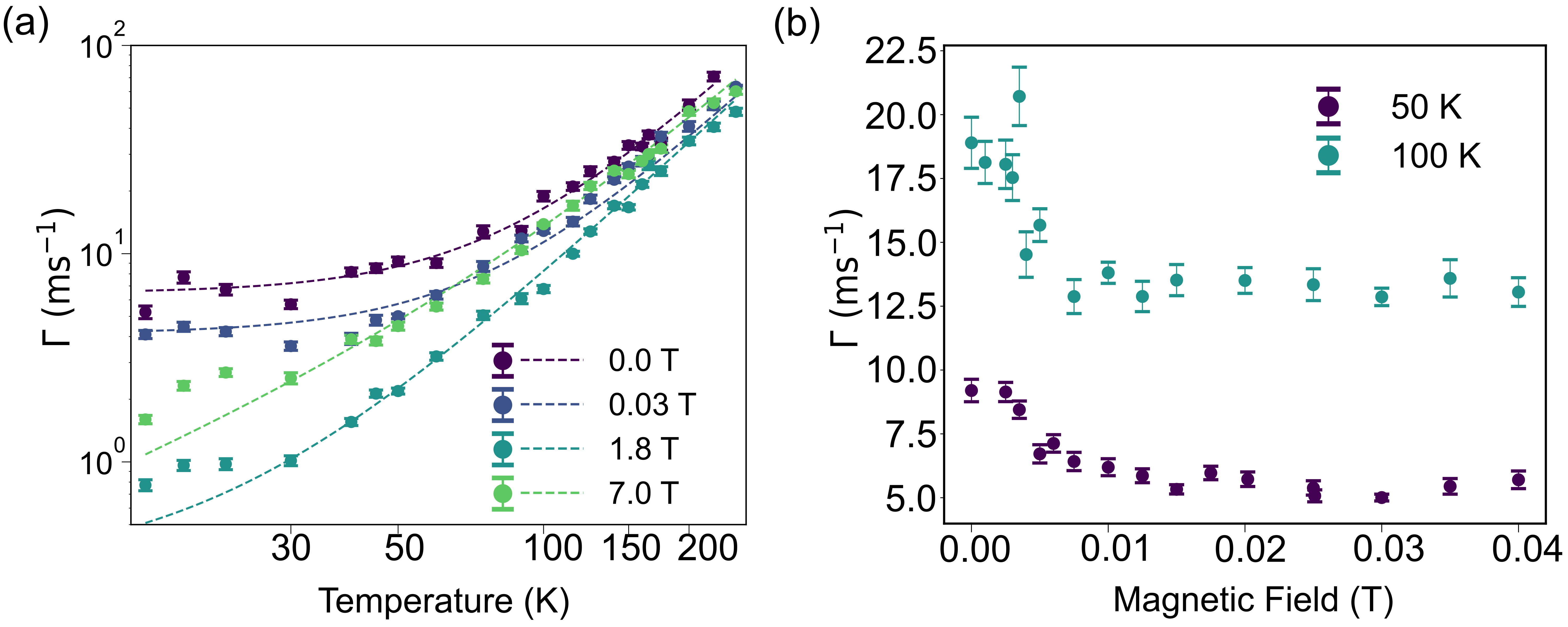}
    \caption{Temperature and magnetic field dependence of the spin relaxation rate of $\mathrm{V_B^-}$ centers in $\mathrm{hBN}$ (a) Temperature dependence of the spin relaxation rate ($\Gamma = 1/T_1$) measured at fixed external magnetic fields $H_{{ext}} = 0$, $0.03$, $1.8$, and $7$ T. At high temperatures ($T > 125$ K), all curves follow a $T^2$ dependence consistent with a two-phonon Raman-like process. In contrast, the low-temperature behavior exhibits strong field dependence, highlighting the presence of additional relaxation channels. Experimental data are overlaid with analytical model fits. (b) Magnetic field dependence of the relaxation rate for fixed sample temperatures $T = 50$ and $100$ K. A sharp suppression of $\Gamma$ is observed below $0.01$ T, attributed to the suppression of resonant spin-spin cross-relaxation, followed by saturation at higher fields.}
    \label{fig:2}
\end{figure}

In Fig.~\ref{fig:2}(a), we present the experimentally measured relaxation rate ($\Gamma=\frac{1}{T_1}$) as a function of temperature for several magnetic fields: $H_{ext}=0, 0.03, 1.8, 7$ T, along with analytical model fits. At temperatures above $125$ K, all curves exhibit a $T^2$ behavior, characteristic of the two-phonon process \cite{hBNT1}. In contrast, at low temperatures, $\Gamma$ depends strongly on magnetic field. For low magnetic fields ($0.03$ T), $\Gamma$ saturates to a constant value at low temperatures, indicative of a weakly temperature-dependent relaxation mechanism. At an intermediate field ($1.8$ T), $\Gamma$ settles at a significantly lower value at low temperatures, suggesting a pronounced magnetic field dependence of the dominant relaxation mechanism. At $7$ T, $\Gamma$ shows a qualitatively different behavior, indicating the emergence of yet another magnetic-field-dependent relaxation mechanism. While prior studies focused on the temperature dependence of $\Gamma$ at low fields \cite{hBNT1}, our measurements reveal a richer dependence of $\Gamma$ on magnetic field across the $0-7$ T range, underscoring the need for a comprehensive study of magnetic-field-dependent relaxation mechanisms.

To this end, we performed magnetic-field-dependent measurement of $\Gamma$ at several fixed temperatures and variable magnetic fields ($H_{ext}$) between $0-7$ T. We first focus on the low-field regime, shown in Fig.~\ref{fig:2}(b), which presents data for $H_{ext} \leq 0.04$ T and temperature $T = 50, 100$ K. In this range, $\Gamma$ decreases sharply between $0-0.01$ T and then plateaus to a field-independent value. The rapid suppression of $\Gamma$ in this field range can be attributed to the reduction of cross relaxation between proximal defects of the same species \cite{jarmola2012}. At zero or low magnetic fields, the $m_s = 0 \leftrightarrow \pm 1$ transitions are close in energy, facilitating nearly resonant cross relaxation between different proximal defects. Applying a modest magnetic field sufficiently separates the energy levels, resulting in a twofold reduction in the relaxation rate \cite{jarmola2012}. This reduction in $\Gamma$ could be leveraged to realize an all-optical quantum sensor that is capable of operating at zero or near-zero external magnetic fields \cite{CrossrelaxationMary}. 

\begin{figure}
    \centering
    \includegraphics[width=\linewidth]{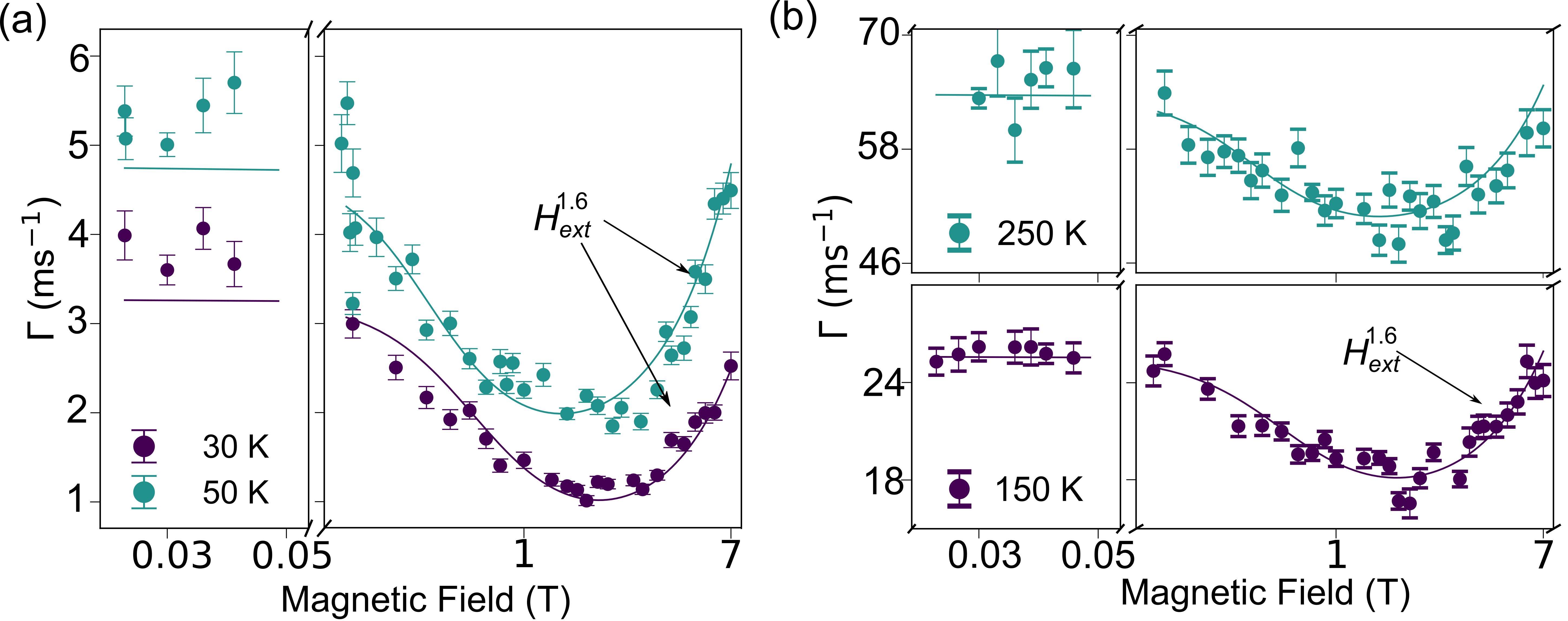}
    \caption{Magnetic-field dependence of the spin relaxation rate of the $\mathrm{V_B^-}$ centers in $\mathrm{hBN}$ (a, b) Magnetic field dependence of the relaxation rate ($\Gamma=\frac{1}{T_1}$) measured at multiple fixed temperatures $T$ = $30$, $50$, $150$, and $250$ K. Data points represent experimental measurements, and the solid lines show analytical model fits. At low magnetic fields ($0.025<H_{{ext}} < 0.045$ T), $\Gamma$ remains nearly constant, followed by a suppression as the field is increased to $1.8$ T. At higher fields ($> 1.8$ T), $\Gamma$ increases approximately as $T\cdot H_{{ext}}^{1.6}$, consistent with a direct spin-phonon relaxation process. Panels are grouped by temperature range for clarity.}
    \label{fig:3}
\end{figure}

To gain further insight, we focus on measurements of $\Gamma$ up to $H_{ext}= 7$ T at several temperatures $T = 30, 50, 150, 250$ K, presented in Fig.~\ref{fig:3}(a, b) along with analytical model fits. The relaxation-rate exhibits a non-monotonic dependence on magnetic field, with a more pronounced relative change at lower temperatures and a minimum close to $1.8$ T. The initial decrease in $\Gamma$ follows a Lorentzian dependence, reminiscent of spin-relaxation of a central spin caused by a rapidly fluctuating spin-bath \cite{surfacespins}. Beyond $1.8$ T, $\Gamma$ increases up to $7$ T, a trend consistent with our previous observation in Fig.~\ref{fig:2}(a). Furthermore, the rate of increase of $\Gamma$ beyond the lowest point is steeper at elevated temperatures, indicating a role of thermal energy at higher field values. The relaxation rate scales with a characteristic $\sim T\cdot{H_{ext}}^{1.6}$ dependence in the high-field regime. Intuitively, this result points to a resonant or 'direct' process scaled by the number of thermal excitations present in the system, such as spin-relaxation caused by a single phonon. In this scenario, the magnetic field scaling can arise from additional phonon density of states resonant with the ground-state splitting at higher fields \cite{maze_1}.

\begin{figure}[H] 
    \centering
    \includegraphics[width=\linewidth]{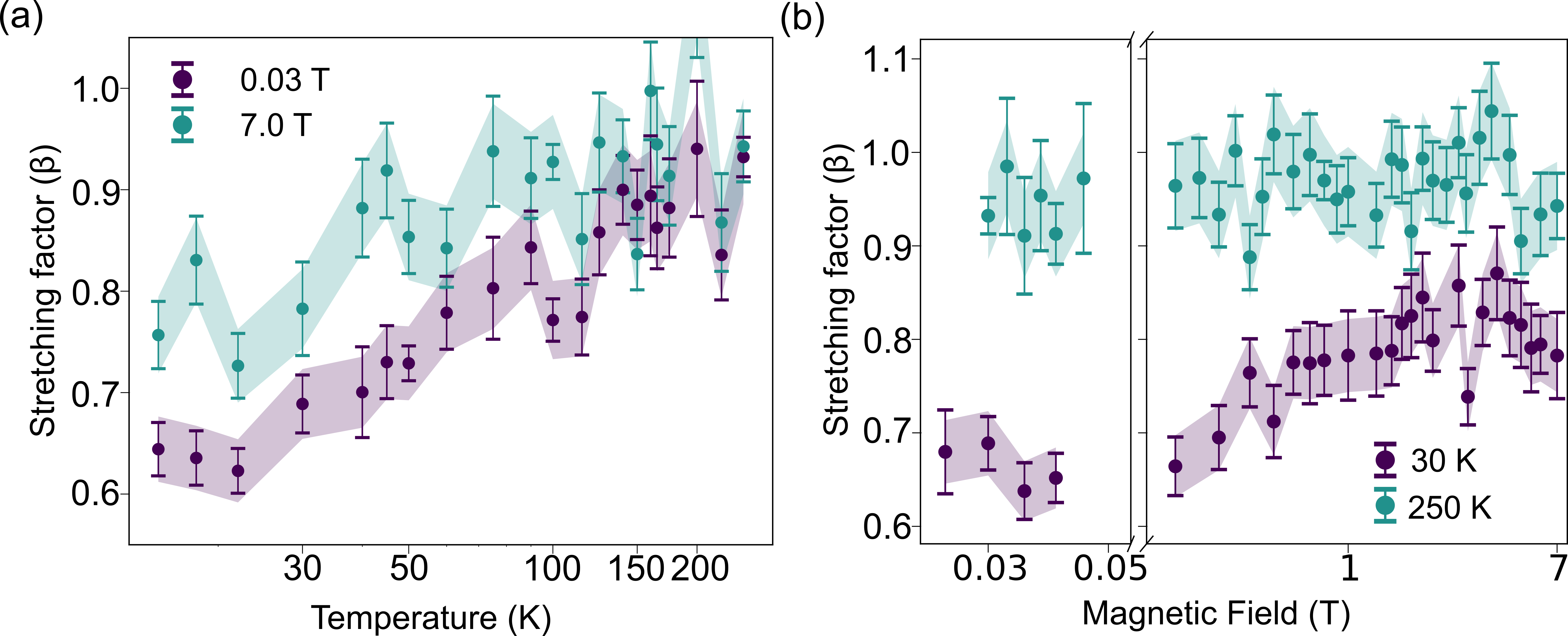}
    \caption{Temperature and magnetic-field dependence of stretching factor ($\beta$) (a) Temperature dependence of the stretching factor $\beta$ for fixed external magnetic fields $H_{ext}=0.03, 7$ T. (b) Magnetic field dependence of $\beta$ at fixed sample temperatures $T = 30$ K and $250$ K, highlighting a pronounced field sensitivity at low temperature and near-constant behavior at high temperature. The evolution of $\beta$ captures the influence of disorder-induced relaxation processes beyond single-exponential dynamics. Shaded bands represent a fixed $\pm5\%$ range around the data points and are used here solely for visualization.}
    \label{fig:4}

\end{figure}

To further elucidate the nature of the spin-relaxation process, we focus on the temperature- and magnetic-field-dependent evolution of the stretching factor ($\beta$). In a disordered environment, a stretched exponential decay ($\beta<1$) reflects a broad distribution of relaxation environments, in contrast to a homogeneous case, where $\beta \sim 1$. Although each spin in the ensemble may decay exponentially, the ensemble-averaged signal follows a stretched exponential profile ($P(t)=exp[-(\frac{t}{T_1})^\beta]$), with the extent of disorder characterized by $\beta$. In Fig.~\ref{fig:4}(a), we present measurements of $\beta$ versus temperature for $H_{ext} = 0.03, 7$ T and complementary measurements of $\beta$ versus magnetic field for temperatures $ T=30, 250$ K in Fig.~\ref{fig:4}(b). Across all fields, $\beta$ consistently decreases as temperature is lowered, with a more pronounced decrease at lower magnetic fields. At low temperatures, $\beta$ increases monotonically with magnetic field, in contrast to the non-monotonic field dependence of the relaxation rate. 

Spin-relaxation in a disordered environment can be influenced by a range of microscopic mechanisms, including fluctuating paramagnetic spins, isotope-disorder, charge dynamics, etc. If spin relaxation is governed solely by local magnetic noise from paramagnetic spins, the stretching factor is $\beta=d_{spin}/2\alpha$, where $d_{spin}$ is the dimensionality of the dipolar spin-spin interactions and $\alpha$ characterizes the long-range magnetic dipolar interaction power law, yielding $\beta=0.5$ for a three-dimensional ensemble \cite{Depolarization, Criticalthermalization, Davis2023}. Additionally, isotope variation and local strain can create inhomogeneities in the spin-phonon coupling, further contributing to nonuniform spin-relaxation dynamics across the measurement area. Moreover, our all-optical measurement scheme may introduce unintended charge dynamics via laser excitation \cite{giricharge, barbosacharge}, leading to a multi-exponential relaxation mimicking stretched exponential decay. To assess the role of charge dynamics, we performed differential measurements at low magnetic fields using a $\pi$-pulse inversion scheme to cancel out non-spin-related contributions. Notably, these measurements also exhibit stretched exponential profiles, suggesting that charge dynamics are unlikely to be the primary source of the observed stretched exponential behavior (see Supplementary Materials). Nonetheless, we note that stretched exponential dynamics cannot be definitively attributed to a single origin without further measurements. 

\subsection*{Phenomenological Description}

Taken together, the dependence of $\Gamma$ and $\beta$ on magnetic field points to a crossover in relaxation dynamics as the field approaches $\sim 1.8$ T. Motivated by these observations, we propose a potential phenomenological model that captures the experimental trends, beginning with spin-phonon interactions. 

In our experiments, the ground-state splitting ($\hbar\omega_0 \sim D \pm g\mu_{B}H_{ext}$) spans $0-0.2$ THz across the measurement range ($0-7$ T), which corresponds to the linear branch of acoustic phonons in the host material $\mathrm{hBN}$. Within this energy range, first-order spin-phonon relaxation is mediated by resonant absorption and emission of phonons ($\omega_{ph}=\omega_0$) \cite{maze_1}. In contrast, second-order spin-phonon relaxation involves non-resonant Raman-like scattering of two higher-energy phonons ($\omega_{ph1}, \omega_{ph2}$), whose energy difference satisfies $\omega_{ph1} - \omega_{ph2} = \omega_0$, resulting in a spin-flip transition \cite{maze_1}. Within a Debye-like model, the combined relaxation rate due to spin-phonon coupling can be expressed as: 

\begin{equation}
    \Gamma^{\textrm{spin-ph}} = A_1\cdot T \cdot {\omega_{0}^{n_1}} + A_2 \cdot T^{n_2},
    \label{spin-ph1}
\end{equation}
where $A_{1}, A_2$ are coupling constants, $T$ is the sample temperature, and $n_1, n_2$ are scaling parameters \cite{jarmola2012,maze_1,maze_2, maze_3,hBNT1}. 

The first term is motivated by the first-order spin-phonon relaxation process. The linear $T$ dependence accounts for the thermal distribution of phonon modes (in the limit $\hbar \omega_0 < k_BT$) resonant with $\omega_0$. For low-energy acoustic phonons, the effective density of states of the phonon-bath and the strength of the linear spin-phonon coupling are modeled by a scaling law ($\omega_{0}^{n_1}$) \cite{maze_1}. This analytical expression reproduces the observed increase in $\Gamma$ following its minimum near $1.8$ T  (for $n_1>0$), highlighting that first-order spin-phonon coupling becomes critical in our case, where large external magnetic fields push $\omega_0$ into a regime where resonant processes are significantly enhanced. We note that deviations from a simple Debye-like model can yield more complex scaling behavior across the frequency range, requiring first-principle investigation \cite{Lunghi2019PhononsRelax}.

The second term accounts for the second-order spin-phonon relaxation process. Previous studies combining first-principle calculations and experiments at low magnetic fields have identified phonon modes near $18$ meV as the dominant contributors to the second-order relaxation process \cite{hBNT1}. These phonon modes are substantially higher in energy compared to the typical ground state splitting of $\mathrm{V_B^-}$ centers \cite{maze_1,hBNT1}. In contrast to the first-order term, the second-order spin-phonon coupling is modeled to be independent of the magnetic field due to its non-resonant nature and depends solely on temperature ($T^{n_2}$) \cite{maze_1, maze_2, maze_3}. 

Next, we model the initial decrease in $\Gamma$ as the field is increased to $1.8$ T. As mentioned before, this feature is empirically described by a Lorentzian noise model. In previous studies, this mathematical framework has been used to describe the coupling of a central spin with a spin bath consisting of randomly fluctuating spins with a characteristic correlation time ($\tau_c$) \cite{surfacespins}. The effectiveness of this mechanism depends on the spectral overlap between the ground state splitting ($\omega_0$) and the magnetic noise power spectrum of the spin bath ($S(\omega)$). The resulting relaxation rate takes the form
\begin{equation}
\Gamma^{\text{Lorentzian}} = \gamma^2 S(\omega_0) = \frac{\eta  \cdot \tau_c}{1 + (\omega_0 \tau_c)^2},
\label{lorentzian}
\end{equation}
where $\gamma$ is the gyromagnetic ratio and $\eta(T)$ is a phenomenological constant that captures the density and dipolar coupling strength of the fluctuating spins. The parameters $\eta$ and $\tau_c$ serve as fit parameters, providing insight into the spin bath properties. 

\begin{figure}[htbp] 
\centering
\includegraphics[width=\linewidth]{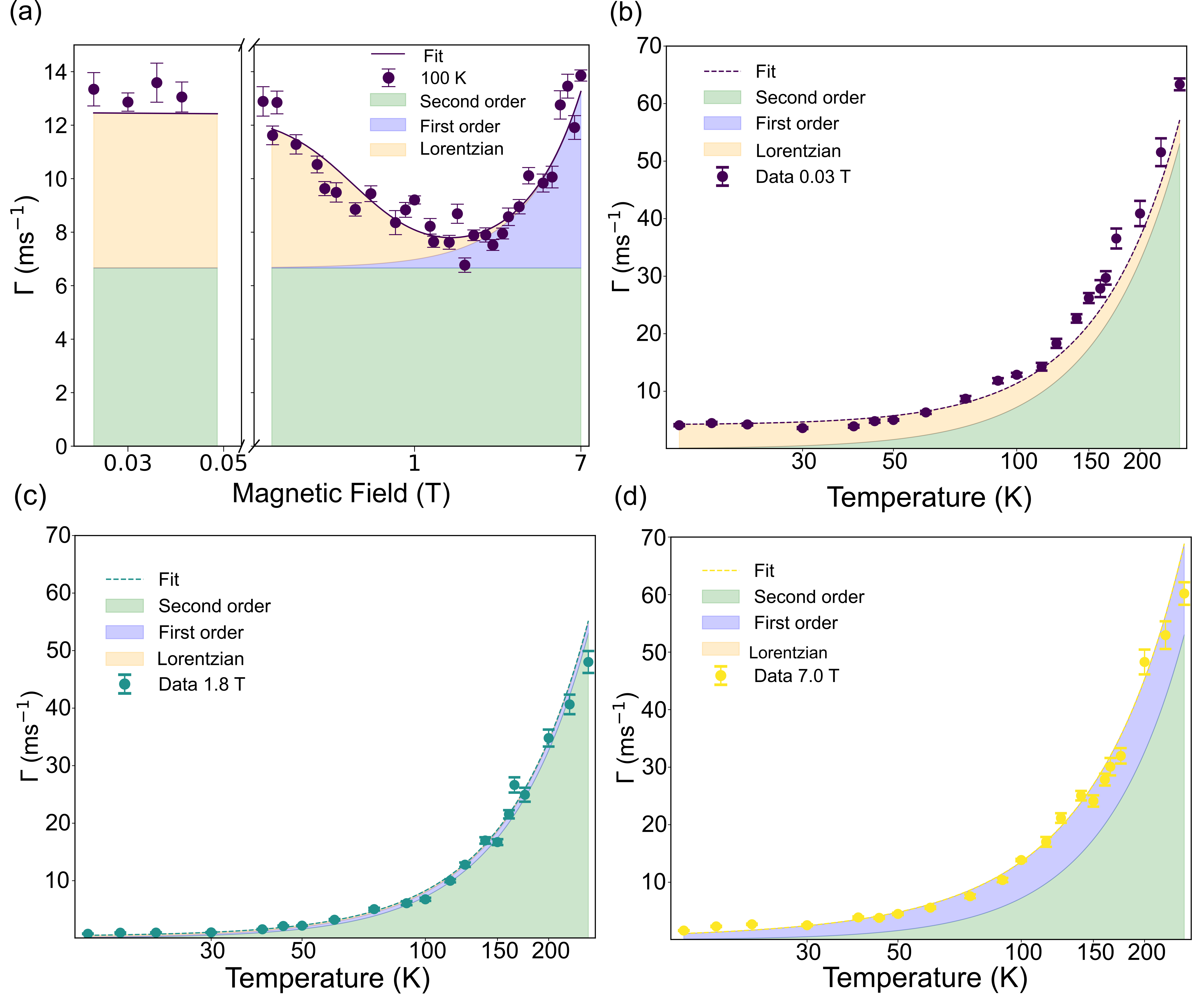}
\caption{Temperature and magnetic field dependence of the spin relaxation rate ($\Gamma = 1/T_1$) of $\mathrm{V_B^-}$ centers in $\mathrm{hBN}$, along with contributions from individual relaxation mechanisms. (a) Magnetic field dependence of $\Gamma$ at $T = 100$ K, showing non-monotonic behavior with a minimum near $1.8$ T. (b–d) Temperature-dependent relaxation rates at selected magnetic fields: (b) $0.03$ T, (c) $1.8$ T, and (d) $7$ T. Experimental data are overlaid with model fits (dashed lines), decomposed into contributions from second-order spin-phonon (green), first-order spin-phonon (purple), and Lorentzian (orange) processes. Stacked shaded regions correspond to the same individual mechanisms as in (b–d). The consistent modeling across both temperature and field dependencies highlights the interplay of multiple relaxation channels in determining the total spin relaxation rate.}
\label{fig:5}
\end{figure}

Based on the physical picture outlined above, we fit the magnetic field-dependent and temperature-dependent relaxation rate ($\Gamma$) with a combined analytical model $\Gamma (H_{ext}, T)=\Gamma^{\textrm{Lorentzian}}+\Gamma^{\textrm{spin-ph}}$ to estimate the relative contributions of the different relaxation mechanisms. A representative fit of the magnetic field dependent data at $T=100$ K, along with contributions from all mechanisms, is presented in Fig.~\ref{fig:5}(a). At low fields, the relaxation is dominated by $\Gamma^{\textrm{Lorentzian}}$. Beyond $1.8$ T, the field-dependent first-order spin-phonon relaxation term grows in magnitude, leading to an observable rise in $\Gamma$. In contrast, the second-order spin-phonon term manifests as a constant background contribution at a fixed temperature. At elevated temperatures, the contribution from the second-order term is higher, thereby reducing the relative variation of $\Gamma$ with $H_{ext}$. Notably, this fitting procedure yields an effective exponent  $n_1\sim 1.6$, implying that at higher magnetic fields, where $\omega_0\propto  H_{ext}$, the effective first-order spin-phonon relaxation rate scales as $\sim H_{ext}^{1.6}$.

To further validate our analytical model, we present temperature-dependent fits, along with the individual contributions from all mechanisms at fixed magnetic fields $H_{ext} = 0.03$, $1.8$, and $7$ T in Fig.~\ref{fig:5}(b–d). These fits reveal similar trends: at low magnetic fields, the relaxation is governed by the Lorentzian noise source, while at higher fields and elevated temperatures, spin-phonon processes become dominant. This analysis yields an exponent $n_2\sim 2$ for all magnetic fields, in agreement with previously observed $T^2$ scaling for experiments at low fields \cite{hBNT1}. The analytical fits obtained from our model show excellent agreement with the experimental trends in the entire parameter space. The extracted parameters are consistent between fits to magnetic field and temperature-dependent data, lending further support to the reliability of our analytical framework. 

While the present model highlights the dominant role of single-phonon processes at high fields, a complete microscopic understanding of spin–phonon interactions requires going beyond the simple Debye-like framework. For instance, the observed non-monotonicity may reflect scaling behaviors beyond those captured in Eq.~(\ref{spin-ph1}), potentially arising from more complex phonon spectral functions \cite{Lunghi2019PhononsRelax}, as well as contributions from flexural phonon modes, interlayer breathing modes, and their coupling to the substrate. In addition, relaxation channels largely independent of temperature and magnetic field, such as dipolar interactions between resonant $\mathrm{V_B^-}$ defects or charge dynamics, may also contribute to the measured rates. These factors complicate the quantitative extraction of coupling constants and bath properties from the present data. A comprehensive microscopic understanding will require first-principles modeling and targeted experiments, such as thickness and doping dependence, which lie beyond the scope of the present study.

\subsection*{Conclusion and Outlook}

In conclusion, utilizing persistent spin-contrast at $7$ T field, we demonstrate that the relaxation of $\mathrm{V_B^-}$ spin defects can be measured from GHz to sub-THz frequencies. Exploiting this functionality, we probe relaxation mechanisms intrinsic to $\mathrm{V_B^-}$ defects across a broad frequency and temperature range. The longitudinal relaxation rate of $\mathrm{V_B^-}$ centers exhibits a pronounced dependence on external magnetic field, highlighting the complex interplay between multiple relaxation mechanisms. While first-order spin-phonon interactions have traditionally been considered important only at extremely low temperatures, our results reveal the crucial role they play at elevated temperatures up to $100$ K at higher magnetic fields. The measured relaxation profiles also exhibit a stretched exponential decay, indicative of disorder in the system.    

Our experiments open exciting possibilities for designing relaxation-based quantum sensors capable of operating at fields as high as $7$ T, corresponding to an energy scale of $\sim 0.2$ THz. Accessing this sub-THz regime is considered challenging due to a lack of reliable sources and detectors \cite{THzgap}. In this context, the all-optical measurement scheme employed in our experiments offers a promising route to probe sub-THz phenomena, such as magnetic modes in antiferromagnets \cite{Antiferromagneticspintronics, NCAntiferromagneticspintronics}. High-field magnetometry enabled by such sensors can be used to investigate exotic condensed-matter phases at several Teslas \cite{arnabscience}. Furthermore, these experiments can be extended to even higher magnetic fields for $\mathrm{V_B^-}$ defects and other spin-defect species, especially those in $\mathrm{hBN}$\cite{Stern2022, Stern2024, tongcangnature}. More broadly, these findings lay the foundation for exploring many-body dynamics in dense ensembles of $\mathrm{V_B^-}$ defects. They also provide key insights into the fundamental relaxation and decoherence mechanisms of $\mathrm{V_B^-}$ defects at high magnetic fields, a crucial step in tailoring these defects for quantum applications.

% If your text is very short, you might need to uncomment the following line to avoid
% layout problems with the figures and tables.
%\newpage

%%%%%%%%%%%%%%%% REFERENCES %%%%%%%%%%%%%%%

\clearpage % Clear all remaining figures and tables then start a new page

% The list of references goes after the main text and before the acknowledgements
% When preparing an initial submission, we recommend you use BibTeX, like this:
%
%\bibliography{main} % for a file named science_template.bib
%\bibliographystyle{sciencemag}

%%%%%%%%%%%%%%%% ACKNOWLEDGEMENTS %%%%%%%%%%%%%%%

\section*{Acknowledgments}

We acknowledge Prof. Norman Yao and Dr. Ashwin Boddeti for useful discussions. We also acknowledge Karthik Pagadala for help with atomic force microscopy. 

\paragraph*{Funding:}
P.U. acknowledges NSF Award No. 1944635 for theoretical support. V.M.S. acknowledges support from an Energy Frontier Research Center funded by the U.S. Department of Energy (DOE), Office of Science, Basic Energy Sciences (BES), under award DE-SC0025620 for cryogenic measurements at Purdue University. B. L acknowledges support from the U.S. Department of Energy, Office of Science, Basic Energy Sciences, Materials Sciences and Engineering Division for cryogenic measurements at Oak Ridge National Laboratory. T.L. acknowledges support by the Gordon and Betty Moore Foundation, grant DOI 10.37807/GBMF12259.

\paragraph*{Author contributions:} 
A.B.S., V.M.S., B.L., and P.U. conceived the idea and planned the experiments. A.B.S. and H.A. performed relaxometry measurements at Purdue University. Y.W., I.G., and B.L. designed, constructed, and performed relaxometry measurements at Oak Ridge National Laboratory. X.G. and T.L. performed the ion implantation of hexagonal Boron Nitride flakes to create the defects. O.M.M. fabricated the microwave waveguide. A.B.S. and Y.P.C. conducted the pick-up and transfer of hBN flakes. A.B.S., H.A., D.S., and A.L. constructed the cryogenic optical experimental setup at Purdue University.  A.B.S., P.A., A.S., Y.P.C., and P.U. analyzed the experimental relaxometry data. A.B.S. wrote the manuscript with support from V.M.S., B.L., P.U., and contributions from all authors. V.M.S., B.L., and P.U. supervised the project. 

\paragraph*{Competing interests:}

There are no competing interests to declare.
%\paragraph*{Data and materials availability:}

% %%%%%%%%%%%%%%%% SUPPLEMENT LIST %%%%%%%%%%%%%%%

% % List the contents of your Supplementary Materials, including the numbers of any
% % supplementary figures, tables, external data files etc. and any references that are
% % cited only in the supplement. In this example, refs. 7-8 are cited only in the supplement.
% % Fill out your numbers accordingly and delete any lines that aren't applicable.
\subsection*{Supplementary materials}
Materials and Methods\\
Supplementary Text\\
Figs. S1 to S10\\
% automatically fills out the last reference number

% %%%%%%%%%%%%%%%% END OF MAIN TEXT %%%%%%%%%%%%%%%

\newpage

% %%%%%%%%%%%%%%%% START OF SUPPLEMENT %%%%%%%%%%%%%%%

% % Figures, tables, equations and pages in the supplement are numbered S1, S2 etc.
\renewcommand{\thefigure}{S\arabic{figure}}
\renewcommand{\thetable}{S\arabic{table}}
\renewcommand{\theequation}{S\arabic{equation}}
\renewcommand{\thepage}{S\arabic{page}}
\setcounter{figure}{0}
\setcounter{table}{0}
\setcounter{equation}{0}
\setcounter{page}{1} % not 0 as \newpage already started a supplementary page
% % References continue the numbering from the main text.

% %%%%%%%%%%%%%%%% SUPPLEMENT TITLE PAGE %%%%%%%%%%%%%%%

\begin{center}
\section*{Supplementary Materials for\\ \scititle}

\author{
Abhishek~Bharatbhai~Solanki$^{1}$, Yueh-Chun~Wu$^{2}$, Hamza~Ather$^{3,4}$,
Priyo~Adhikary$^{1}$, \and Aravindh~Shankar$^{1}$, Ian~Gallagher$^{2}$, Xingyu~Gao$^{3}$, Owen~M. Matthiessen$^{1,4}$, \and
Demid~Sychev$^{1,4}$, Alexei~Lagoutchev$^{4}$, Tongcang~Li$^{1,3,4,5}$, Yong~P.~Chen$^{1,3,4,5,6,7}$, \and
Vladimir~M.~Shalaev$^{1,3,4,5,\ast}$, Benjamin~Lawrie$^{2,\ast}$, Pramey~Upadhyaya$^{1,5,\ast}$ \and

\small$^\ast$Corresponding authors. Emails:  prameyup@purdue.edu, lawriebj@ornl.gov, shalaev@purdue.edu
}
\end{center}

% % Fill out the numbers for each type of supplementary material,
% % and delete any lines that aren't applicable.
% % These are just example numbers that don't match the rest of this template.
\subsubsection*{This PDF file includes:}
Materials and Methods\\
Supplementary Text\\
Figures S1 to S10\\

\newpage

% %%%%%%%%%%%%%%%% MATERIALS AND METHODS %%%%%%%%%%%%%%%

\subsection*{Materials and Methods}

A single crystal of hexagonal Boron Nitride ($\mathrm{hBN}$) with naturally abundant nuclear isotope distribution was mechanically exfoliated onto a standard $\mathrm{SiO_2/Si}$ substrate. The entire substrate, containing flakes of varying thicknesses, was implanted using helium ions, at a dosage of $1 ~\mathrm{ion/nm^2}$ at 2.8 keV. Suitable $\mathrm{hBN}$ flakes were identified using an optical microscope and transferred to a fresh $\mathrm{Si}$ substrate to avoid background fluorescence from the damaged $\mathrm{Si}$ substrate, using standard pick-up and transfer procedure. A suitable $\mathrm{hBN}$ flake ($\sim 42$ nm in thickness) was identified post-transfer and characterized with room-temperature optical measurements.  

Cryogenic experiments at Purdue University and Oak Ridge National Laboratory were performed using an Opticool cryostat from Quantum Design. The cryostat is equipped with a $7$ T superconducting magnet, with the magnetic field aligned along the out-of-plane direction relative to the sample. Optical measurements were performed with a $532$ nm laser from Hubner Photonics. The excitation beam was incident normal to the sample surface, with both excitation and collection performed through a $100 x$ in-vacuum objective. The excitation beam was scanned across the sample using a 4f system with a galvoscanner. The resultant photoluminescence (PL) signal was detected using a Si-based single-photon-avalanche-detector (SPAD) from Thorlabs. The PL signal was filtered with a $700 $ nm long-pass filter to selectively detect emission from $\mathrm{V_B^-}$ defects. The laser beam was modulated by an acousto-optic modulator (AOM) to generate $\sim 10 \mu s$ laser pulses for spin initialization and readout. The SPAD output was recorded via a gated counter in a National Instruments (NI) data acquisition card (DAQ), with a counter window duration of $1.5 \mu s$. A pulse streamer 8/2 from Swabian Instruments was used to synchronize the experiment. 

Microwave measurements for a differential measurement scheme, aimed at eliminating charge dynamics, were performed using a gold waveguide ($300$ nm thick) patterned onto a sapphire substrate. An alumina spacer layer ($\sim 750$ nm thick) was deposited on top of the gold-waveguide to minimize coupling of Johnson noise emanating from the gold waveguide to the ground state of the $\mathrm{V_B^-}$ defects. A suitable $\mathrm{hBN}$ flake ($\sim 50$ nm thick) was transferred onto the alumina spacer layer. The microwave waveguide was designed using COMSOL simulations to maximize the coupling between the microwave mode with the $\mathrm{V_B^-}$ defects. Microwave pulses were generated using a gated microwave switch. 

% \subsubsection*{Example supplement heading}

% The two main sections of the supplement can be split up using headings.

% %%%%%%%%%%%%%%%% SUPPLEMENTARY TEXT %%%%%%%%%%%%%%%

\subsection*{Supplementary Text}

\subsubsection*{Flake Characterization}

\begin{figure}[H]
\centering
\includegraphics[width=0.5\linewidth]{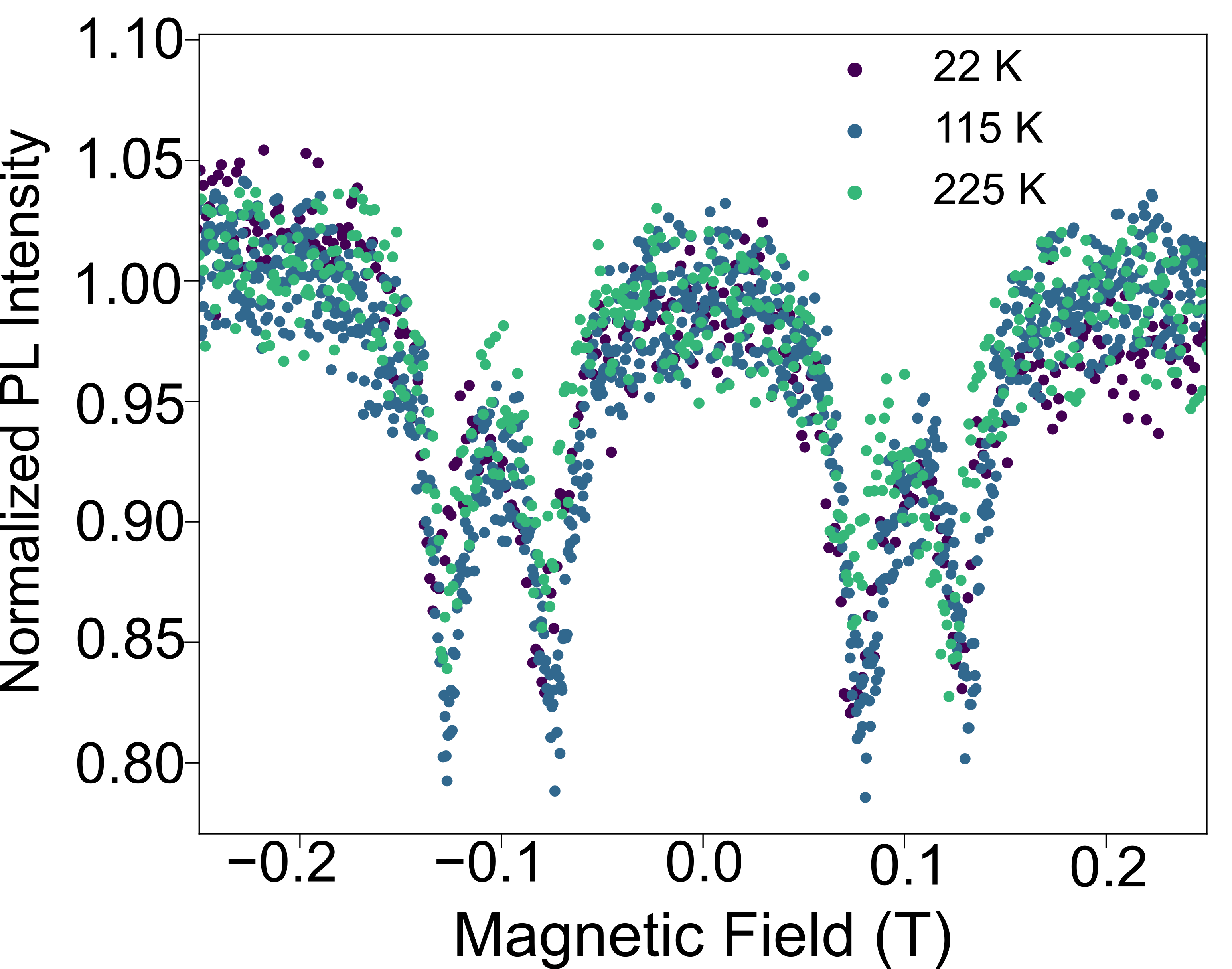}
\caption{Normalized Photoluminescence signal (PL) strength vs magnetic field.}
\label{fig:PL_vs_field}
\end{figure}

Fig.~\ref{fig:PL_vs_field} presents the photoluminescence (PL) intensity as a function of external magnetic field ($H_{ext}$) for the $\mathrm{hBN}$ flake containing uniformly implanted $\mathrm{V_B^-}$ defect centers, measured at temperatures of $T = 22, 115,$ and $225$ K. Pronounced PL dips are observed at $H_{ext} \sim \pm 0.075$ T and $\pm 0.125$ T, corresponding to level anti-crossings in the excited and ground states, respectively \cite{Jacques-field-dependenthBN}. In our experiments, we focus on magnetic fields $H_{ext}< 0.05$ T, $H_{ext} > 0.17$ T, to avoid the level anti-crossings, where the spin dynamics become more complex and out of the scope of this work.

\begin{figure}
    \centering
    \includegraphics[width=0.5\linewidth]{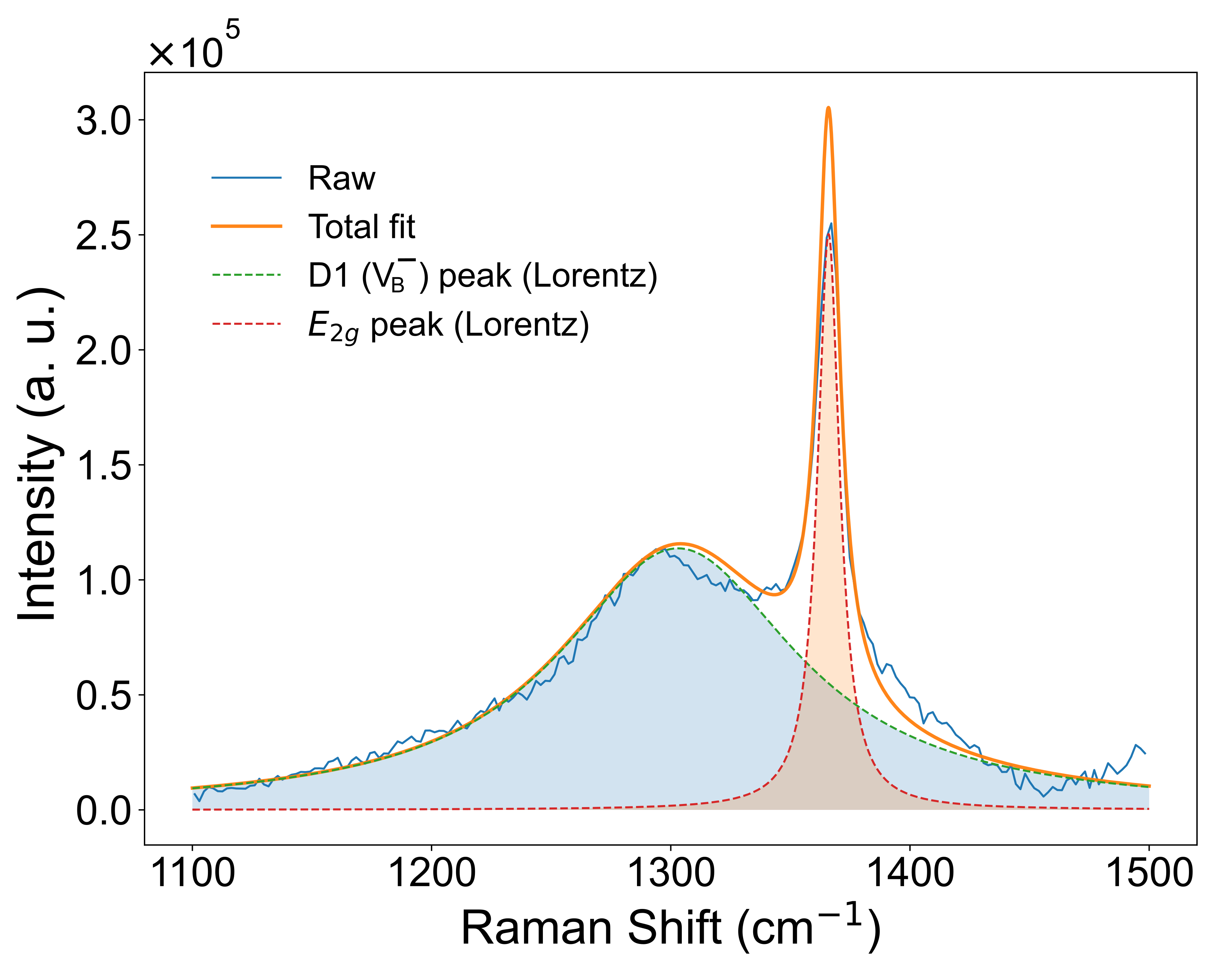}
    \caption{Raman scattering measurement of the $\mathrm{hBN}$ flake employed in this work with uniformly implanted $\mathrm{V_B^-}$ centers. }
    \label{fig:Raman}
\end{figure}

Figure~\ref{fig:Raman} shows the Raman spectrum of the $\mathrm{hBN}$ flake containing uniformly implanted $\mathrm{V_B^-}$ defect centers, measured under ambient conditions using a commercial Raman imaging microscope (Thermo Scientific DXR3xi) with a $532$ nm excitation source and a $50x$ objective. The spectrum is well described by two Lorentzian peaks in the range $1100$–$1500~\mathrm{cm^{-1}}$. The higher-energy $E_{2g}$ mode at $\sim 1380~\mathrm{cm^{-1}}$ is intrinsic to $\mathrm{hBN}$ and remains unaffected by implantation. In contrast, the lower-energy $D_1$ peak at $\sim 1300~\mathrm{cm^{-1}}$ emerges upon implantation and is attributed to $\mathrm{V_B^-}$ centers, with its intensity reflecting the defect density.  

\begin{figure}[H]
    \centering
    \includegraphics[width=0.75\linewidth]{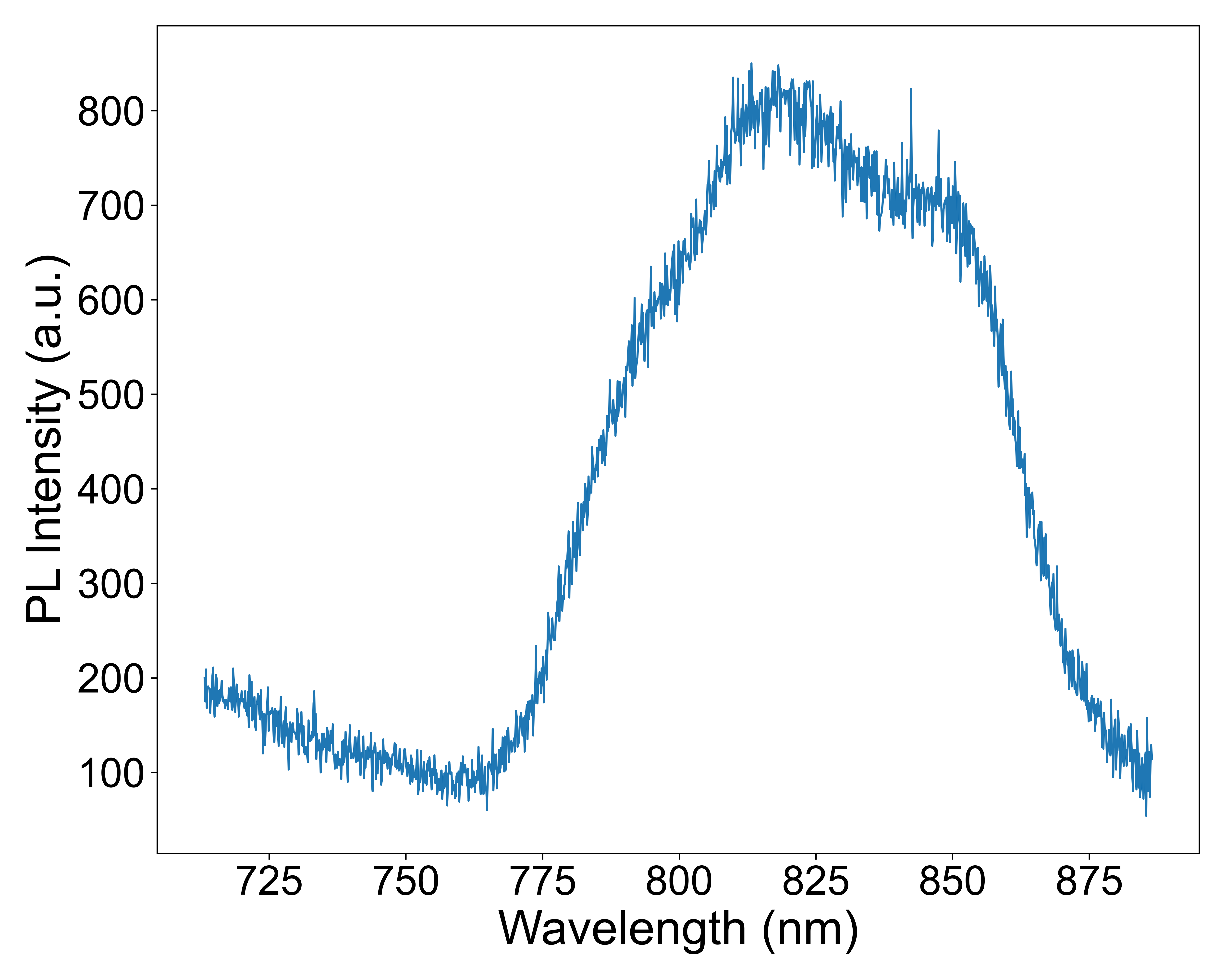}
    \caption{Photoluminescence spectrum of $\mathrm{V_B^-}$ centers in $\mathrm{hBN}$ at temperature $T=3$ K.}
    \label{fig:spectra}
\end{figure}

Fig.~\ref{fig:spectra} presents the photoluminescence spectrum of $\mathrm{V_B^-}$ centers in $\mathrm{hBN}$ measured at temperature $T=3$ K with a $700$ nm long-pass filter. 

\begin{figure}[H]
    \centering
    \includegraphics[width=\linewidth]{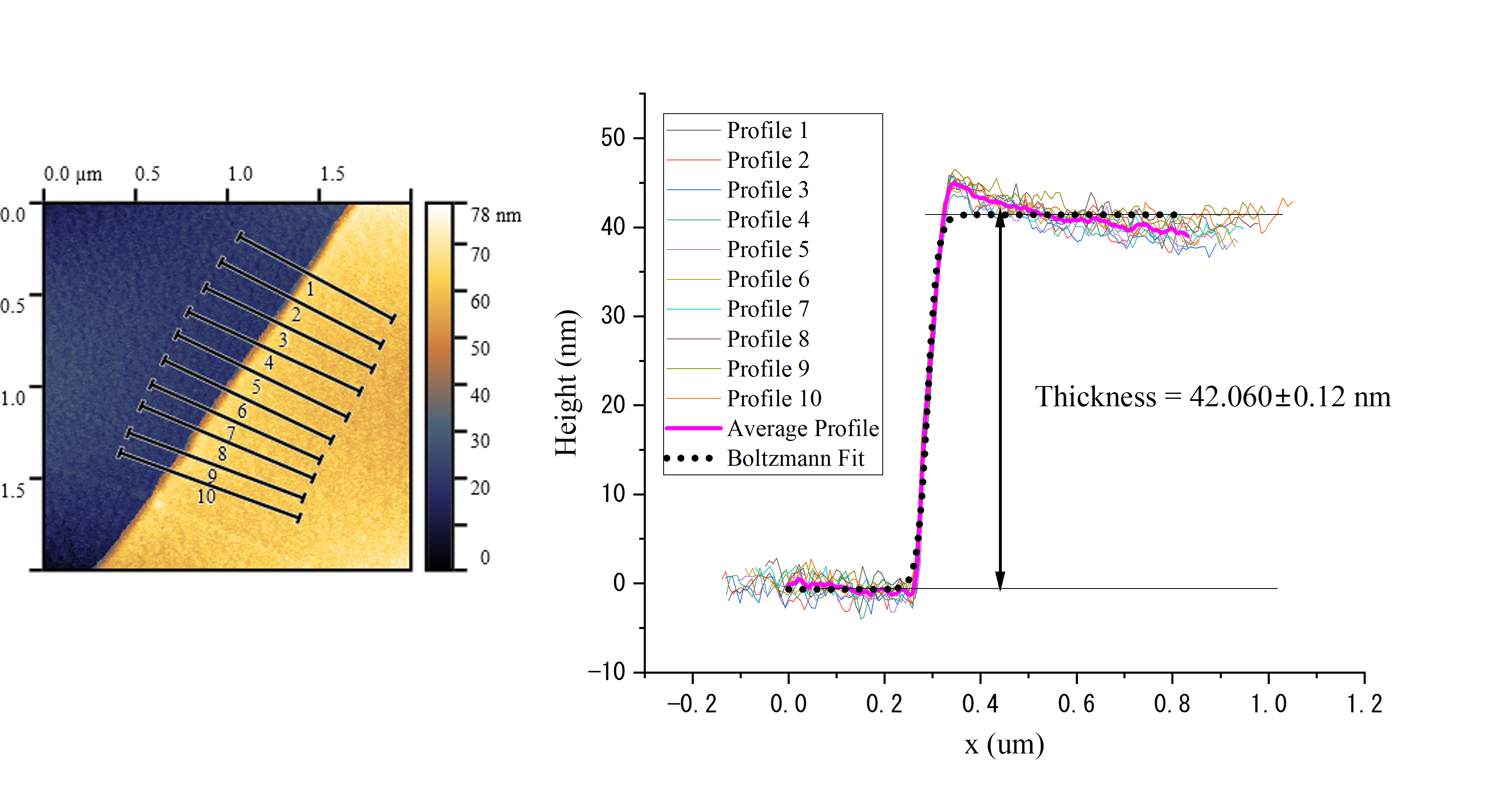}
    \caption{Atomic Force Microscopy image of the $\mathrm{hBN}$ flake on $\mathrm{SiO_2/Si}$ substrate. The average thickness of the flake is $\sim 42$ nm.}
    \label{fig:AFM}
\end{figure}

Fig.~\ref{fig:AFM} presents the atomic force microscopy (AFM) image of the $\mathrm{hBN}$ flake employed in this work for the measurements presented in the main text. The flake exhibits a uniform thickness of $\sim 42$ nm. 

\subsubsection*{Analytical fits of the experimental relaxation data with a stretched exponential}

\begin{figure}
    \centering
    \includegraphics[width=\linewidth]{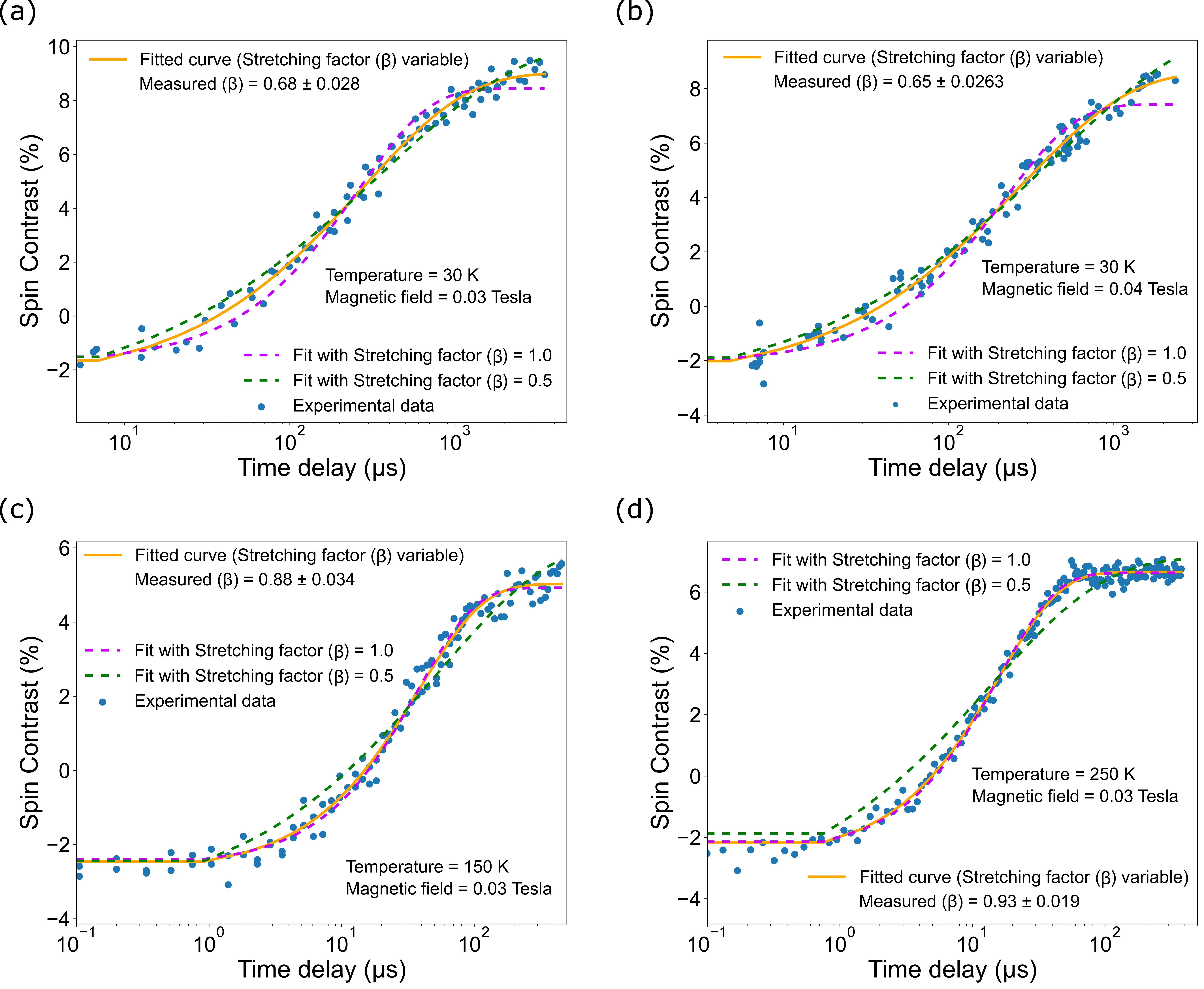}
    \caption{Measured $T_{1}$ relaxation curves (blue dots) of $\mathrm{V_B^-}$ centers in $\mathrm{hBN}$ at selected magnetic fields and temperatures, overlaid with fits to a stretched-exponential relaxation model (solid orange line). Additional fits with stretching factors fixed at $\beta = 1$ and $\beta = 0.5 $ are included for comparison. (a, b) Magnetic field = $0.03 , 0.04$ T and Temperature $T = 30$ K. (c) Magnetic field = $0.03$ T and Temperature $T = 150$ K. (d) Magnetic field = $0.03$ T and Temperature $T = 250$ K.}
    \label{fig:Stretched_exponential}
\end{figure}

As mentioned in the main text, the spin-contrast as a function of delay time ($\Delta t$) is well described by a stretched exponential of the form $C_0(1-\exp{{(-(\frac{\Delta t}{T_1})^\beta)}})$ \cite{Depolarization}. Here, $C_0$ is the spin-contrast amplitude, $T_1$ is the spin relaxation time, and $\beta$ is the stretching factor. These parameters are extracted by fitting the experimental data at various temperatures and magnetic fields. Representative fits are shown in Fig.~\ref{fig:Stretched_exponential}: (a,b) $H_{ext} = 0.03, 0.04$ T at $T = 30$ K; (c) $H_{ext} = 0.03$ T at $T = 150$ K; and (d) $H_{ext} = 0.03$ T at $T = 250$ K. For comparison, fits using fixed stretching exponents of $\beta = 0.5$ and $\beta = 1$ are also included. The deviation of these fixed-exponent fits from the experimental data underscores the importance of treating $\beta$ as a free parameter alongside $\Gamma = \frac{1}{T_1}$ during the fitting process.  

To further investigate the origin of the stretched exponential decay behavior, we performed laser-power-dependent measurements for excitation powers over the range $0.8$ to $4.0$ mW. The experimental dataset, overlaid with stretched exponential fits, is presented in Fig.~\ref{fig:laser_power_dependence}. The measurement does not reveal a significant dependence of the measured relaxation rate or the stretching factor on the excitation power. On the other hand, the measured spin-contrast ($C_0$) exhibits a clear power dependence as illustrated in Fig.~\ref{fig:laser_power_dependence}. Across all measured powers, a consistent offset of $-2.5 \%$ is present. While the precise origin of this offset, whether arising from charge dynamics or experimental artifacts, remains uncertain, its magnitude is considerably smaller than the measured spin-contrast ($C_0 > 10\%$). It is not expected to affect the extracted parameters significantly.

\begin{figure}[H]
	\centering
	\includegraphics[width=\textwidth]{Combinedlaserpowerdependence.png} 
    \caption{Laser power dependence of the relaxation rate ($\Gamma$) and stretching factor ($\beta$) at temperature $T = 30$ K and magnetic field $H_{ext} = 0.02$ T. The observed variation in the measured relaxation rate and stretching factor lie within the fitting uncertainties. Panels (a-e) correspond to laser powers of 0.8, 1.5, 2.2, 3.1, 4.0 mW, respectively.}
    \label{fig:laser_power_dependence}
\end{figure}

To further evaluate the robustness of the fitting procedure, we refitted the experimental data after intentionally excluding data corresponding to dark time values ($\Delta t$) below a cutoff time. We present the result of this analysis for temperatures $T = 30, 50$ K in Fig.~\ref{fig:datamask_30K} and Fig.~\ref{fig:datamask_50K} respectively. In panels (a,b) of both figures, the experimental relaxation curves measured at $H_{ext}=0.03, 7$ T are overlaid with fits obtained by systematically varying the cutoff time between $0.01$ to $20~\mu s$. Panels (c,d) display the corresponding magnetic-field dependence of the extracted relaxation rate $\Gamma$ and stretching factor $\beta$ as a function of the cutoff time. No systematic dependence on the cutoff time is observed, aside from isolated deviations at a few points. These results reinforce the robustness of the fitting approach and confirm that the initial negative offset does not significantly affect the extracted parameters.

\begin{figure}[H]
	\centering
	\includegraphics[width=\textwidth]{Supplementary_image_datamask_30K.png} 
    \caption{Robustness of the stretched-exponential fitting procedure at $T = 30$ K. Panels (a,b) show representative relaxation curves measured at $H_{ext} = 0.03$ and $7$ T, overlaid with fits obtained by excluding data below cutoff times ranging from $0.01$ to $20~\mu{s}$. Panels (c,d) display the extracted relaxation rate $\Gamma$ and stretching exponent $\beta$ as a function of magnetic field for the same range of cutoff times. No systematic dependence on the cutoff is observed, aside from isolated deviations at a few points.}
    \label{fig:datamask_30K} 
\end{figure}

\begin{figure}[H]
	\centering
	\includegraphics[width=\textwidth]{Supplementary_image_datamask_50K.png} 
    \caption{Robustness of the stretched-exponential fitting procedure at $T = 50$ K. Panels (a,b) show representative relaxation curves measured at $H_{ext} = 0.03$ and $7$ T, overlaid with fits obtained by excluding data below cutoff times ranging from $0.01$ to $20~\mu{s}$. Panels (c,d) display the extracted relaxation rate $\Gamma$ and stretching exponent $\beta$ as a function of magnetic field for the same range of cutoff times. No systematic dependence on the cutoff is observed, aside from isolated deviations at a few points.}
    \label{fig:datamask_50K} 
\end{figure}

\subsubsection*{Microwave assisted differential measurements}

In our experiments, we use $10~\mu s$ pulses of $532$ nm laser to initialize and readout of the ground-state spin population distribution of the defects. Initialization relies on spin-dependent intersystem crossing (ISC) that polarizes the ground state into the $m_s=0$ spin sublevel. A variable dark interval ($\Delta t$) then allows the system to relax toward thermal equilibrium, after which a second pulse reads out the population via the same spin-dependent pathway. Ideally, the laser pulses would only modify the population distribution within the ground-state spin manifold. However, the excitation laser pulse can also alter the population of the negatively charged boron vacancy ($\mathrm{V_B^-}$) centers, via photo-ionization and recombination processes, as observed for $\mathrm{NV}$ centers in diamond \cite{giricharge, barbosacharge}. This transient charge distribution also relaxes towards the steady-state charge distribution during the dark interval. The transient dynamics of $\mathrm{V_B^-}$ population modulates the $\mathrm{PL}$ signal strength. Consequently, the measured relaxation profile reflects a sum of both spin- and charge-relaxation processes. 

\begin{equation}
\label{eq:spin-charge-model}
S(\Delta t)
= A_{\mathrm{s}}\!\left[1 - e^{-(\Delta t/T_{1})^{\beta}}\right]
+ A_{\mathrm{c}}\!\left[1 - e^{-\Delta t/T_{{c}}}\right]
+ C \, .
\end{equation}

\noindent\textbf{Parameters.}
\begin{itemize}
  \item \(\Delta t\): dark interval between pump and readout pulses.
  \item \(A_{\mathrm{s}}\): spin-contrast amplitude.
  \item \(T_{1}\): longitudinal spin-relaxation time.
  \item \(\beta \in (0,1]\): stretching exponent related to spin-relaxation
  \item \(A_{\mathrm{c}}\): amplitude due to charge re-equilibration (its sign encodes whether PL rises or falls).
  \item \(T_{{c}}\): charge re-equilibration time constant.
  \item \(C\): baseline offset (e.g., residual background).
\end{itemize}

The spin-relaxation parameters ($A_s, T_1$) are expected to exhibit temperature and magnetic field dependence due to spin-phonon and spin-spin interaction \cite{jarmola2012}, whereas transient charge-dynamics parameters ($A_c, T_c$) should exhibit a much stronger dependence on the wavelength of the excitation laser and excitation power \cite{Depolarization}. If the charge-dynamics rate ($1/T_c$) is indeed independent of temperature and magnetic field, it would manifest as a constant offset in the measured spin-relaxation rates. However, the presence of multiple exponential processes can effectively mimic a stretched exponential profile. Furthermore, the extent of charge dynamics could have a spatial variation, which could further contribute to the observed stretched-exponential profile. To disentangle these different processes, we employ a microwave-assisted differential measurement scheme designed to eliminate transient charge-related dynamics and recover the intrinsic spin-relaxation behavior. Since coplanar waveguides cannot be used to coherently drive the $\mathrm{V_B^-}$ centers at Tesla-scale magnetic fields, due to ground state splitting $>20$ GHz, we restrict the measurements to external magnetic fields of a few hundred Gauss.  

\begin{figure}[H]
	\centering
	\includegraphics[width=\textwidth]{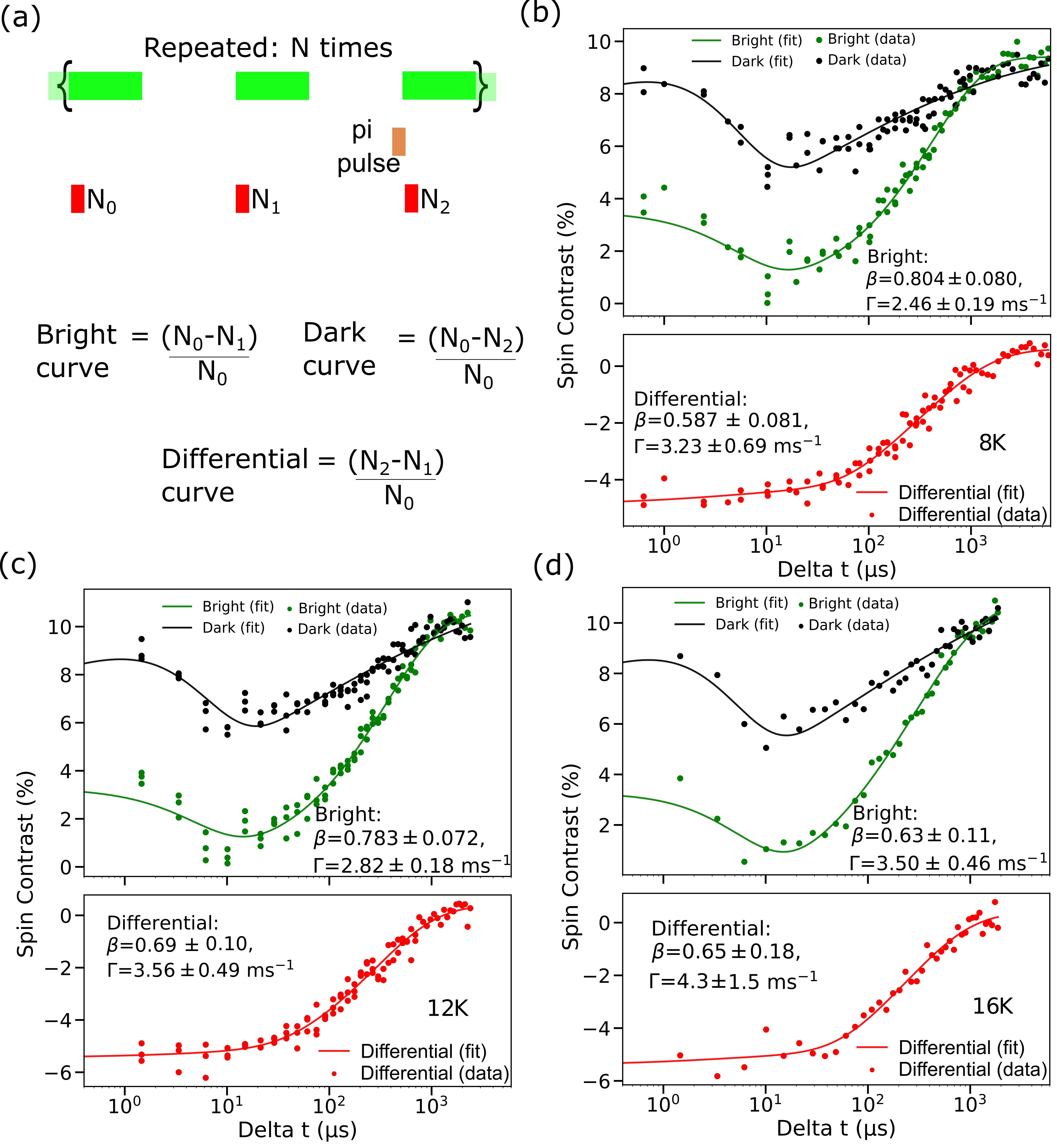} 
	\caption{Microwave-assisted differential measurement scheme to identify charge-dynamics contributions. (a) Schematic of the differential protocol, where alternating microwave on/off sequences are used to isolate spin-related relaxation signals from background charge effects. (b–d) Representative relaxation curves measured at temperature $T = 8, 12, 16$ K, shown with microwave-on (green), microwave-off (black) and differential (red) traces overlaid with fits.}
    \label{fig:pi_pulse} 
\end{figure}

This modified measurement scheme is illustrated in Fig.~\ref{fig:pi_pulse}(a). The first green pulse initializes the ground state into the $m_s=0$ state. After a variable dark interval, a second green pulse is applied to read out the ground-state population. The number of photons collected in a short time window at the onset of this pulse ($N_1$) provides a measure of the time-dependent spin population. Toward the end of the same pulse, the spin is reinitialized into the $m_s = 0$ state, followed by another dark interval of equal duration. A resonant $\pi$-pulse (duration $\sim 10$ ns) is then applied before a third green pulse is used for readout. The photons detected in the corresponding short interval at the beginning of this pulse yield $N_2$, which reflects the spin population after the $\pi$-pulse. The number of photons measured at the beginning of the first green pulse serves as a reference signal ($N_0$) corresponding to the ground state spin initialized into the $m_s=0$ state. 

The bright curve, defined as $\frac{N_0-N_1}{N_0}$, corresponds to the microwave-free all-optical relaxation signal used for the measurements presented in the main text. The dark curve is measured as $\frac{N_0-N_2}{N_0}$ and the differential curve is measured as $\frac{N_2-N_1}{N_0}$. The dark curve carries the same laser pulse-induced charge dynamics contribution as the bright curve ($A_{\mathrm{c}}\!\left[1 - e^{-\Delta t/T_{{c}}}\right]$), which is effectively canceled out in the differential scheme. 

In Fig.~\ref{fig:pi_pulse}(b,c,d) we present the measurements performed at temperatures $T=8, 12$ and $16$ K, respectively. The results show that the differential curves also exhibit a stretched exponential decay profile. Interestingly, the stretching in the differential curve is even more pronounced than in the bright curve. The differential scheme yields a relaxation rate $\Gamma$ that is marginally higher than the value obtained from the bright curve; however, the increased fitting error reflects the smaller signal contrast inherent to this approach. This observation suggests that transient charge dynamics make only a minor contribution in the all-optical measurement scheme, which would otherwise result in a reduced rate in the differential scheme.

In addition to these observations, we note that the significant magnetic field dependence of the relaxation rate presented in the main text naturally rules out the possibility of charge dynamics dominating the measurement scheme. Intuitively, we expect the extent of charge dynamics to be dependent on the energy levels of the conduction band and valence band of electrons relative to the energy levels of the negatively-charged and neutral $\mathrm{V_B^-}$ centers. These energy scales would determine the ionization rate for a given wavelength of excitation laser. However, these rates are not affected by the magnetic field, which only leads to a change in the ground state spin splitting.

We also note that the differential scheme is only effective at rejecting the pulse-synchronous, laser-induced charge dynamics which result in a non-equilibrium charge distribution. As discussed above, this transient charge state relaxes towards a steady-state charge state which could result in an additional exponential curve ($A_{\mathrm{c}}\!\left[1 - e^{-\Delta t/T_{{c}}}\right]$). This additional exponential is rejected in microwave assisted differential scheme. However, it doesn't rule out the possibility of steady-state charge dynamics after the charge state population has stabilized during the dark interval ($\Delta t >> T_{{c}}$) in the measurement scheme. In a real system, steady-state population exchange between different charge species is expected to occur ($\mathrm{V_B^-} \Leftrightarrow \mathrm{V_B^0}$). In this scenario, the population of $\mathrm{V_B^-}$ is fixed, resulting in a fixed $\mathrm{PL}$ signal. But the charge dynamics would result in a loss of spin-information. Therefore, charge dynamics could manifest as an additional channel for spin-relaxation. These complex factors can be disentangled by a systematic investigation of the relaxation rate for different implantation dosages and different excitation laser wavelengths, which is beyond the scope of this work. 

\subsubsection*{Data analysis and fitting}

As outlined in the Phenomenological Description section of the main text, we model the magnetic-field and temperature dependence of the relaxation rate using a combined analytical framework. 

The spin–phonon contribution is described by

\begin{equation}
\Gamma^{\textrm{spin-ph}} = A_1T\omega_{0}^{n_1} + A_2T^{n_2},
\label{spin-ph2}
\end{equation}

where $A_1$ and $A_2$ are coupling constants, $T$ is the sample temperature, and $n_1$ and $n_2$ are scaling exponents \cite{jarmola2012,maze_1,maze_2,maze_3,hBNT1}.

The remaining contribution is captured by a Lorentzian form,

\begin{equation}
\Gamma^{\text{Lorentzian}} = \gamma^2 S(\omega_0) = \frac{\eta\tau_c}{1 + (\omega_0 \tau_c)^2},
\end{equation}

where $\gamma$ is the gyromagnetic ratio, $\eta(T)$ is a phenomenological parameter reflecting the density and dipolar coupling of fluctuating spins, and $\tau_c$ is the correlation time. Here, $\eta$ and $\tau_c$ serve as fitting parameters, providing quantitative insight into the properties of the spin bath. 

The total relaxation rate is then expressed as a sum of the two contributions, 

\begin{equation}
    \Gamma (H_{ext}, T)=\Gamma^{\textrm{Lorentzian}}+\Gamma^{\textrm{spin-ph}}, 
\end{equation}

which allows us to extract the relative contributions of the different relaxation mechanisms. 

An important distinction between the two processes is that the double quantum transition between $m_s=+1 \leftrightarrow m_s=-1$ spin-levels can be driven by spin–phonon coupling, whereas such a process is not permitted through spin–spin interactions \cite{maze_1}. In our experiments, the splitting between the $m_s=+1 \leftrightarrow m_s=-1$ states reaches as high as $392$ GHz at $H_{ext}=7$ T, which can have important implications for the first-order spin-phonon coupling. Therefore, the most general form of spin-phonon contribution can be expressed as

\begin{equation}
\Gamma^{\textrm{spin-ph}} = A_1T(\frac{D + g\mu_{B}H_{ext}}{\hbar})^{n_1} +  A_1T(\frac{D - g\mu_{B}H_{ext}}{\hbar})^{n_1} +  A_1T(\frac{ 2g\mu_{B}H_{ext}}{\hbar})^{n_1} + A_2T^{n_2},
\label{spin-ph3}
\end{equation}

In this expression, we assume that the coupling constant $A_1$ and the exponent for the first-order spin phonon coupling $n_1$ remain the same for the $m_s=0 \leftrightarrow m_s=\pm1$ and $m_s=+1 \leftrightarrow m_s=-1$ transitions \cite{maze_1}. This assumption can be tested by first-principles calculations and further measurements. For $H_{ext}>>\frac{D}{g\mu_B}$, where the first-order term is expected to be crucial, and $\hbar\omega_0<k_BT$, the first-order spin-phonon coupling is $\sim T(H_{ext})^{n_1}$.

To ensure consistency between the temperature- and field-dependent analyses, we employ an iterative fitting approach. Specifically, the exponent $n_1$ is first estimated from fits to the magnetic-field–dependent data and then fixed in the temperature-dependent analysis to extract $n_2$. The extracted value of $n_2$ is subsequently used as input for refining the magnetic-field fits. This process is repeated until the parameters converge, yielding a mutually consistent set of fit values for $n_1$ and $n_2$.

\subsubsection*{Additional temperature dependent data}

\begin{figure}[H]
	\centering
	\includegraphics[width=\textwidth]{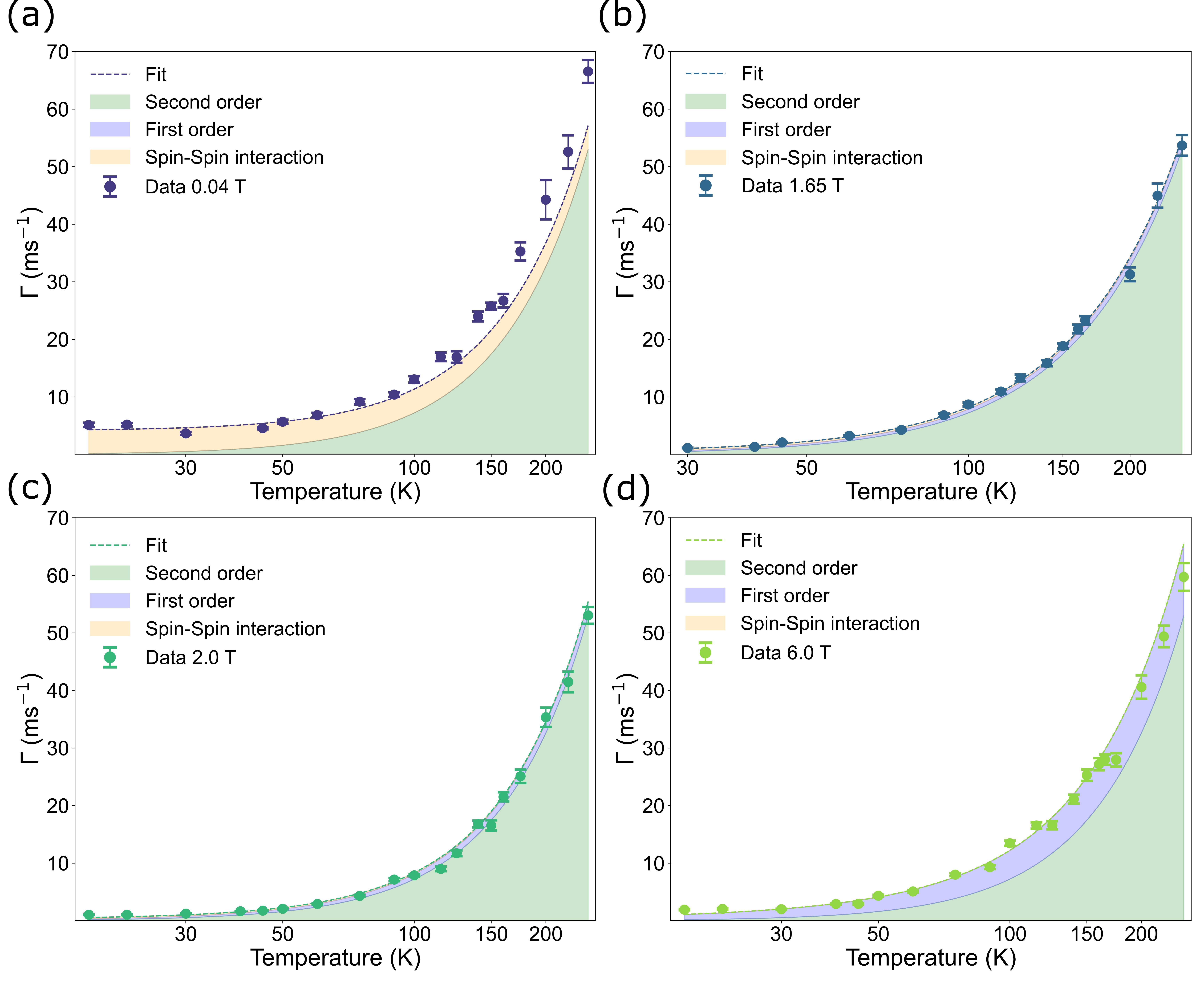}    
    \caption{Temperature dependence of the relaxation rate ($\Gamma=1/T_1$) at selected magnetic fields, showing the decomposition into individual contributions. Panels (a–d) correspond to $H_{ext}=0.04, 1.65, 1.8, 6.0$ T, respectively.}
    \label{fig:invT1_vs_T_combined}
\end{figure}

Additional temperature-dependent data at fixed magnetic fields $H_{ext} = 0.04, 1.65, 2.0, 6.0$ T is presented in Fig.~\ref{fig:invT1_vs_T_combined}. 

% %%%%%%%%%%%%%%%% SUPPLEMENTARY TABLES %%%%%%%%%%%%%%%

% %%%%%%%%%%% CAPTIONS FOR OTHER SUPPLEMENTARY FILES %%%%%%%%%%

% \clearpage % Clear all remaining figures and tables then start a new page

% \paragraph{Caption for Data S1.}
% \textbf{All captions must start with a short bold sentence, acting as a title.}
% Then explain what is included in the supplementary data file.
% Give as much detail as you would for a table e.g. explain the meaning of every column,
% units used, any special notation etc.

% %%%%%%%%%%%%%%%% SUPPLEMENTARY REFERENCES %%%%%%%%%%%%%%%

% % Do NOT include a reference list in the supplement.
% % All references must be in a single list at the end of the main text.
% % The copyeditors will ensure that the correct reference list appears with each version of the paper
% % (print, HTML, PDF, mobile app, metadata for bibliographic databases etc.)

\end{document}